\documentstyle[aas2pp4]{article}
\lefthead{Bergh\"ofer et al.}
\righthead{The Thermal Pressure of the ISM}

\begin{document}

\title{The Thermal Pressure of the Hot Interstellar Medium derived from Cloud 
Shadows in the Extreme Ultraviolet}

\author{Thomas W. Bergh\"ofer}
\affil{Space Sciences Laboratory, University of California, Berkeley, 
	CA 94720, USA}

\author{Stuart Bowyer}
\affil{Space Sciences Laboratory, University of California, Berkeley, 
	CA 94720, USA}

\author{Richard Lieu}
\affil{Department of Physics, University of Alabama, Huntsville AL 35899, USA}

\author{Jens Knude}
\affil{Niels Bohr Institute for Astronomy, Physics, and Geophysics, University
	Observatory, Juliane Mariesvej 30, DK-2100 Copenhagen OE, Denmark}

\begin{abstract}
We have used the Deep Survey telescope of EUVE to investigate shadows
in the diffuse EUV/Soft X-Ray background cast by clouds in the
interstellar medium.  We confirm the existence of a shadow
previously reported, and provide evidence for two new shadows.
We used IRAS data to identify the clouds producing these
shadows and to determine their optical depth to EUV radiation. 
The EUV-absorbing clouds are optically thick in the EUV,
and all EUV emission detected in the direction of these shadows must
be produced from material in front of the clouds.  We obtained
new optical data to determine the distance to these clouds.  
We use a new differential cloud technique to obtain the pressure of the
interstellar medium.
These results do not depend on any zero level calibration of the data.
Our results provide evidence that the pressure of the hot interstellar gas is 
the same in three different directions in the local interstellar medium, and 
is at least 8 times higher than derived for the local cloud 
surrounding our Sun. This provides new evidence for large thermal pressure 
imbalances in the local ISM, and directly contradicts the basic assumption of thermal 
pressure equilibrium used in almost all present models of the interstellar
medium.
\end{abstract}

\keywords{ISM: bubbles, clouds, general, structure --- X-rays: ISM}

\section{Introduction}

The discovery of the soft X-ray / EUV background (\cite{bow68})
and its anti-correlation with the Galactic H{\sc i} distribution has led
to an on-going debate on the origin of that emission. Since an absorption
column density of $N_{\rm H{\sc i}} = 1.2 \times 10^{20}$\,cm$^{-2}$ (which 
corresponds to an optical depth $\tau = 1$ at 0.25\,keV) is reached at a 
distance of approximately 100\,pc in the direction of the Galactic plane, the 
diffuse background must arise from hot gas in the local interstellar medium
(ISM). Optical and UV absorption line measurements of nearby stars together
with observations in the X-ray and EUV wavelength range have 
established a local region with a significant H{\sc i} deficiency which is 
partly filled with a plasma, the so-called ``Local Bubble'' (LB). 
No definite model exists for this region and its origin is unclear. 

One view is that the gas in the LB has been heated by a supernova 
explosion which occurred in the solar neighborhood about $10^7$\,yr ago 
(cf. \cite{mckost77}, \cite{coxrey87}). Other authors claim that the LB is 
part of an asymmetrically shaped superbubble created by stellar winds and 
supernova explosions (cf. \cite{frisch95}). However, the physical properties 
such as density, temperature, pressure, and extension of the LB, are not well
determined and so far observational constraints are insufficient to establish 
a canonical model of the evolution and the origin of the LB. 
The McKee \& Ostriker model is the best known model but it is not easily 
testable. Only soft X-ray and EUV observations provide a direct method
to study the hot gas in the ISM and test LB models.

This paper presents new measurements for the thermal pressure in the local
ISM, using observations of nearby clouds taken by the 
Deep Survey (DS) telescope on-board the {\em Extreme Ultraviolet Explorer} 
(EUVE). Prior to this paper, Bowyer~et~al.~(\cite{bow95}) reported on the first
detection of
a spatial absorption feature in the diffuse EUV background; this feature was 
positionally coincident with an IRAS cirrus cloud at a distance of $\la$
40\,pc. Since the cloud casting this shadow is optically thick 
at EUV wavelengths, the authors concluded that all residual EUV emission 
observed at the
position of the cloud originated from material along the line of sight
in front of the cloud. By attributing the background subtracted flux in front 
of the cloud to the hot gas in the ISM, Bowyer~et~al.~(\cite{bow95}) derived 
for the length 
of the emitting region (40\,pc) a pressure (P/$k$) of 19,000\,cm$^{-3}$K. 
These authors pointed out that an imperfect zero level calibration of the 
data can lead to a smaller value for the pressure, and
estimated a lower limit for the pressure of 7,000\,cm$^{-3}$K. 

The purpose of this paper is to present a re-observation of the previously
reported cloud shadow in the EUV background and observations of two new 
shadows discovered in the DS. For all three shadow regions we provide
new optical photometry data which establishes the distance of the respective 
clouds. We employ a differential cloud technique which does not depend on any 
zero level calibration, to obtain the pressure of the ISM in these directions.

The outline of this paper is as follows: First, in Sect.~2 we describe our
new observations and the data analysis. In Sect.~3, we employ a new
analysis technique and compare our results for the three cloud shadows. We
derive physical quantities of the hot interstellar gas from our observations. 
Finally, in Sect.~4, we discuss the implications for models of the local 
ISM.

\section{Observations and reduction methods}
\label{datared}

\subsection{EUVE observations}
\label{euvobs}

After launch in June 1992 the EUVE satellite performed an all-sky
survey in four bandpasses in the spectral range between 70 and 800\,\AA. 
Details about the EUVE mission and the instrumentation can be found in 
\cite{boma91}. Here we only summarize the 
capabilities of the DS instrument which has been used to obtain
the EUVE data discussed in this paper. The DS instrument, consisting of a
grazing incidence telescope and two broad band filters (Lexan/Boron filter: 
65--190\,\AA, Al/Carbon filter: 160--385\,\AA) in front of an imaging 
microchannel plate detector, is mounted perpendicular to the other 
scanner telescopes. During the EUVE survey phase, the DS made deep
exposures of the sky along the ecliptic plane. A schematic layout of the
DS detector field  can be found in Lieu~et~al.~(\cite{lieu93}). Owing to its 
anti-solar pointing direction, the DS instrument observed about 
$2\arcdeg \times 1\arcdeg$ of the Galactic plane per day. With a typical 
exposure of 20\,ksec/pixel the DS is about a factor of 10 more 
sensitive than the EUVE all-sky survey. 
Generally speaking, diffuse emission is difficult to detect with 
microchannel plate detectors which exhibit a large instrumental background. 
It is thus clear that the DS data are better suited for a search for
diffuse structures in the EUV background than the much shorter exposed EUVE 
all-sky survey data.

Lieu~et~al.~(\cite{lieu93}) showed that the diffuse astrophysical background 
contributes about 25\% of the total count rate observed in the Lex/B filter
of the DS instrument, and used a `filter frame subtraction' method to
remove particle background contamination from the diffuse background 
data. The resulting Lex/B count rates were correlated with such
terrestrial coordinates as satellite position, solar zenith angle and
magnetic L-value, to identify and eliminate periods of data when the
count rates are higher than the stable minimum value.

We first discuss our re-observation of lb165-32. As a naming convention for 
the shadowing clouds discussed here we use the letters ``lb'' followed by the 
Galactic coordinates of the cloud in degrees. In this re-observation
we used a different observational technique than that employed in the original
observation. The longitudinal side of the filter was maintained at a
90$\arcdeg$ orientation with respect to the ecliptic plane to maximize
the exposed region above and below the plane.  In the
original observation during the EUVE DS this filter was 
constantly rotating about the DS axis. 
In both the original DS scan as well as in the re-observation, the 
same field of view of $1.53\arcdeg \times 0.49\arcdeg$, excluding the area of 
a dead spot ($0.093\arcdeg \times 0.069\arcdeg$) was observed.
However, the rotating filter configuration
during the DS leads to a shorter exposure of those parts
of the scan located outside the inner $0.5\arcdeg$ swath centered
on the ecliptic plane. Depending on the size, shape, and orientation of the 
interstellar absorption feature, this could result in a difference between
the DS and re-observation count rates.
We carried out a series of tests with a variety of synthetic absorption 
features folded with the spatial detector response and established a
change in the count rates by up to 20\%.

\begin{figure}
\plotone{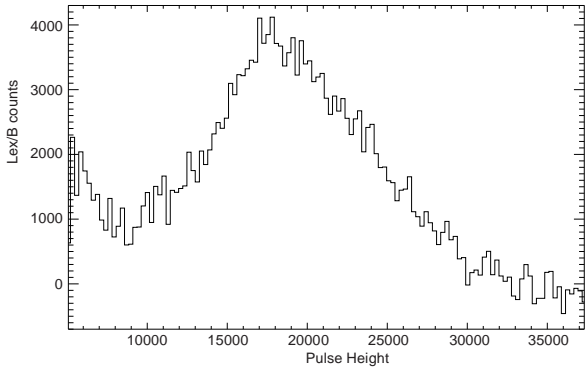}
\caption{\label{euvedata} Pulse height spectrum of the Lex/B filter counts
observed during the re-observation of lb165-32. The large roughly Gaussian 
peak is identical to that seen in observations of EUV point sources showing
that most of the counts registered in this observation are due to photons.}
\end{figure}

In Figure~\ref{euvedata} we plot a pulse height spectrum of the Lex/B filter 
counts detected during the re-observation of the cloud lb165-32.
The observed spectrum presented in Figure~\ref{euvedata} consists of a 
prominent Gaussian profile and an exponential background distribution. 
The Gaussian profile is typical of photon data from point sources and
indicates that more than 90\% of the diffuse counts are contributed by 
photons. This confirms and strengthens the results of 
Lieu~et~al.~(\cite{lieu93}) who showed using different techniques that we
are detecting photons, not charged particles, in these observations. Our new
analysis technique (discussed later), shows that about 50\% of these photons
are astronomical in origin. The nonastronomical photons are believed to be
a combination of scattered geocoronal and solar EUV and X-ray emission.

\begin{figure}
\plotone{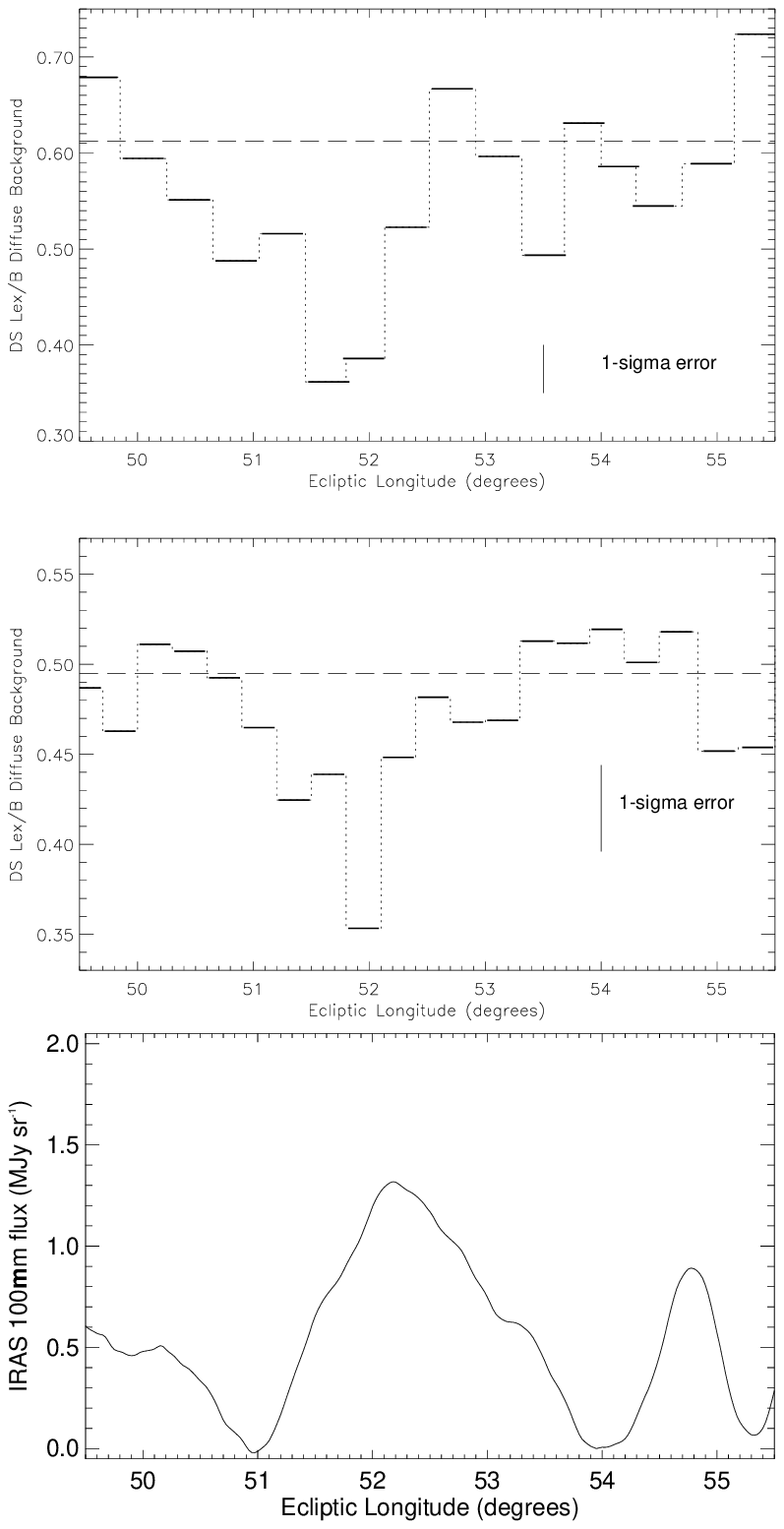}
\caption{\label{165euv} EUV diffuse background count rate in the Lex/B filter
as a function of ecliptic longitude; 
the upper panel shows the new observation of lb165-32, the middle panel 
represents the same section obtained during the Deep Survey. The dashed lines 
correspond to the average background levels (cf. Sect.~\ref{euvobs}).
For comparison, the lower panel shows the IR emission scan constructed from
the continuum subtracted IRAS 100$\mu$m sky maps folded with the
spatial response of the DS Lex/B filter.}
\end{figure}

In the upper panel of Figure~\ref{165euv}, we plot our new results on the 
diffuse background as a function of the ecliptic longitude for lb165-32. 
For comparison, the original observation is shown in the middle panel; the
plotted scan of the ecliptic plane begins at Galactic coordinates 
$l \approx 163.3\arcdeg, b \approx -34.3\arcdeg$ and ends at 
$l \approx 167.5\arcdeg, b \approx -29.5\arcdeg$.
It is evident that an absorption feature is present in both data sets.
The dashed lines (upper and middle panel) give the average background levels 
for the two scans outside the shadow region. In the case of the DS
observation, this background has been determined from the scan range between 
ecliptic longitude $39\arcdeg$ and $60\arcdeg$. For the re-observation we
determined the background from the scan at ecliptic longitude above 
$52.5\arcdeg$ and below $50.3\arcdeg$.
The respective average diffuse background count rates are 
0.495 $\pm$ 0.06\,cts/s for the DS scan and 0.612$\pm$ 0.026\,cts/s 
for the re-observation; the count rates correspond to a field of view 
of 0.7433 square degree. In the lower panel of 
Figure~\ref{165euv} we plot an IR-emission scan across lb165-32 which we 
constructed from the continuum 
subtracted IRAS 100\,$\mu$m sky map that has been folded with the spatial 
response of the rotating DS Lex/B filter region used to extract the EUV 
background scan across lb165-32. As can be seen from Figure~\ref{165euv}, 
the absorption feature in the EUVE data is positionally coincident with an 
enhanced emission feature in the IRAS 100$\mu$m data in the ecliptic plane 
near ecliptic longitude $l = 52\arcdeg$. 
In both independent EUVE observations the absorption feature appears with a 
significance of 3 -- 3.5\,$\sigma$ below the average background 
level. These two independent measurements, both at greater than than 
3\,$\sigma$ leave no doubt about the detection of this shadow 
in the EUV background. 

Modeling of synthetic absorption profiles folded with the detector spatial 
responses shows that the different shape of the absorption feature and the 
roughly 20\% higher count rates during the re-observation of lb165-32 is 
explained by the modified observing technique used in the re-observation.

For the DS scan of lb165-32 we find at the position of the cloud 
a minimum count rate of $0.35 \pm 0.06$\,cts/s. From the scan regions 
immediately adjacent to the cloud ($l = 50.0 - 50.6\arcdeg$ and 
$l = 53.3 - 54.8\arcdeg$) we determine a local continuum diffuse EUV 
background rate of 0.51 $\pm$ 0.06\,cts/s. 

\begin{figure}
\plotone{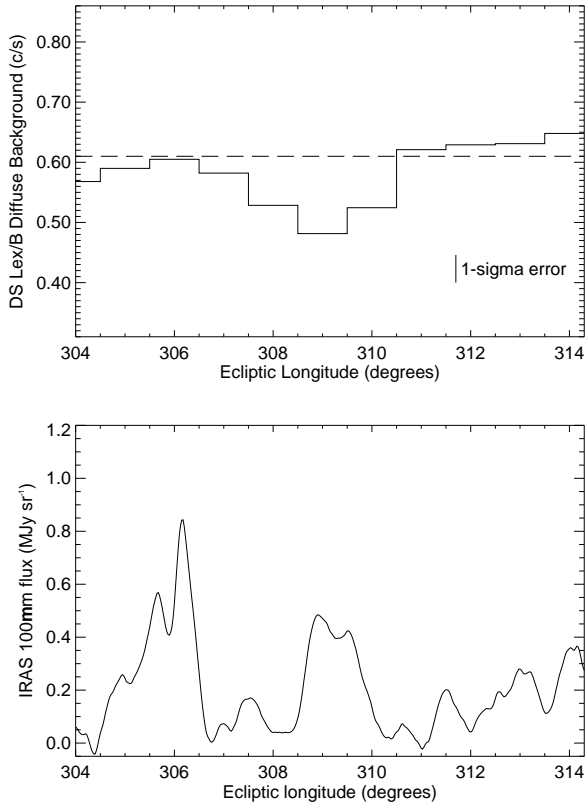}
\caption{\label{27euv} Plot of the EUV background scan across lb27-31 
and the respective IR 100$\mu$m emission scan (cf. middle and lower panel of 
Figure~\ref{165euv}).}
\end{figure}

\begin{figure}
\plotone{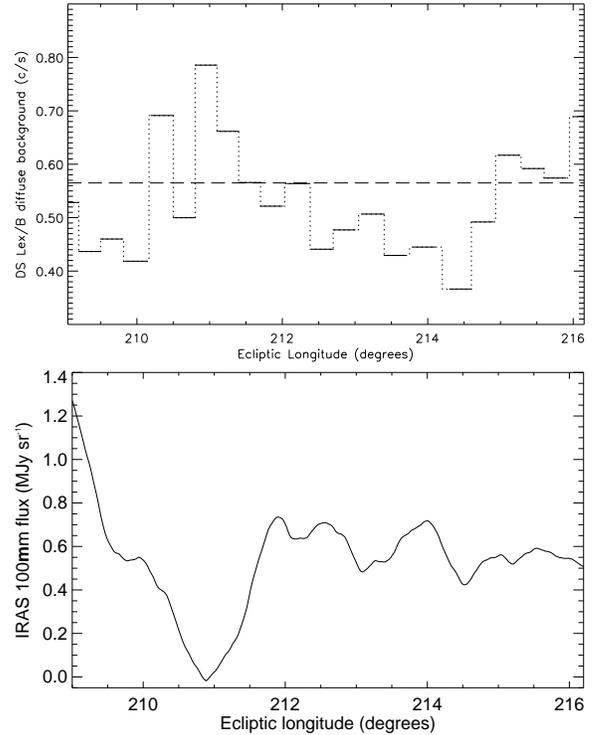}
\caption{\label{329euv} EUV diffuse background count rate in the Lex/B filter
as a function of ecliptic longitude; 
the upper panel shows the new observation of lb329+46, the lower panel shows 
the IR emission scan constructed from the continuum subtracted IRAS 100\,$\mu$m
sky maps folded with the spatial response of the DS Lex/B filter.}
\end{figure}

We have found a second absorption feature in the EUV background which we
designate as lb27-31.
In the upper panel of Figure~\ref{27euv} we show the EUVE DS scan of the 
ecliptic plane near ecliptic longitude $l = 309.5\arcdeg$ beginning at 
Galactic coordinates $l \approx 24.9\arcdeg, b \approx -29.0\arcdeg$ and 
ending at $l \approx 32.0\arcdeg, b \approx -37.0\arcdeg$. The lower panel of
Figure~\ref{27euv} shows the IR emission for this sky location. Both the EUV 
background as well as the IR emission have been reduced in the same way as 
described for lb165-32. 
The EUV background absorption feature in the Galactic plane centered near
ecliptic longitude $l = 309.0\arcdeg$ appears at the position of an IRAS
cirrus cloud. Again, a comparison of the EUVE DS data with the IRAS 100$\mu$m 
data provides evidence for a prominent (5\,$\sigma$) shadow in the EUV 
background cast by a cloud in the local ISM.
The minimum count rate of $0.48 \pm 0.03$\,cts/s at the position of the 
shadowing cloud is taken from the data point at $l = 309\arcdeg$.
By averaging the observed count rates between $l = 305 - 307\arcdeg$ and
$l = 311 - 313\arcdeg$, we determine a local continuum count rate of 
$0.61 \pm 0.03$\,cts/s, 

A third cloud shadow (designated as lb329+46) was found in the DS 
data in the range 
$l = 209 - 216\arcdeg$ (corresponding to Galactic coordinates between 
$l \approx 324.4\arcdeg, b \approx 49.4\arcdeg$ and $l \approx 332.0\arcdeg, 
b \approx 44.6\arcdeg$). This scan is shown in Figure~\ref{329euv}.
With respect to the mean DS Lex/B diffuse background count rate (dashed line 
in Figure~\ref{329euv}), taken from the scan range between 
$l = 205 - 220\arcdeg$ excluding the region $l = 210 - 215\arcdeg$,
the EUV background exhibits a 4\,$\sigma$ absorption 
feature between $l = 212.5 - 215\arcdeg$ as well as an enhanced 
emission feature near $l = 211\arcdeg$. The comparison with the
IRAS 100$\mu$m data shows that the absorption feature is spatially coincident
with two IR emission features near $l = 212.8\arcdeg$ and 
$l = 214\arcdeg$; the IRAS 100\,$\mu$m sky map shows two finger-like
emission structures crossing the ecliptic plane at these positions (note that 
the zero flux of the IR scan is a result of the subtracted continuum 
which has been taken from the region near $l = 211\arcdeg$). 
We adopted a count rate of $0.36 \pm 0.05$\,cts/s for the position of the 
shadow. From the regions near $l = 212\arcdeg$ and 
$l = 215.5\arcdeg$ we obtain a local EUV background count rate of 
$0.57 \pm 0.05$\,cts/s.

The significantly lower IR emission near $l = 211\arcdeg$ indicates a 
much lower absorption column in this direction. The higher EUV emission ($0.70 \pm 0.05$\,cts/s) in this region can be explained by a larger emission 
column along this line of sight.

\subsection{Optical observations}
\label{optobs}

In this subsection we present our optical observations of stars in the 
direction of these shadows. In order to determine the distance to the
EUV-absorbing clouds we have used detailed Str\"omgren photometry; this 
technique has been shown to provide the absolute magnitude and reddening of
most classes of main-sequence stars (cf. \cite{stroem66}, \cite{crawf79}, 
\cite{olsen88}). All photometry was carried out with the Str\"omgren automatic
telescope at ESO La Silla, Chile. The stars for observation were selected from
the position and proper motion star catalog (\cite{roes91}) and the Space 
Telescope guide star catalogue 
(\cite{lask90}). The error propagation of the uvby$\beta$ indices result in a
typical error of roughly 15\% for the distance to the stars investigated.
For 34 of our 210 program stars Hipparcos parallaxes exist. For these stars
the distances obtained by Str\"omgren photometry agree within the errors of the
Hipparcos measurements. This provides confirmation of the accuracy of our 
distance estimates for the overlapping stars. A detailed discussion of this 
comparison including all our optical data of the individual stars in the cloud 
regions can be found in \cite{knude97}. In Table~\ref{stars} we provide a 
summary of the optical data of the stars that we use to establish the distance
of the shadowing clouds. For each of these 8 stars, Table~\ref{stars} provides 
the position in Galactic coordinates, the distance derived from our photometry 
data and from the Hipparcos data, and the reddening E$_{b-y}$.
\begin{table*}
\caption[ ]
{\label{stars} Summary of stellar data constraining the distance of the 
shadowing clouds}
\begin{center}
\begin{tabular}{llccc}
\hline \hline 
Name & gal. Coord. & \multicolumn{2}{c}{Distance (pc)} & E$_{b-y}$ \\
     & l, b & phot. & Hipp. & mag \\
\hline
HD\,20065 & 51.034, 0.298 & 39.0 & 40.8 & 0.029 \\
HD\,20477 & 51.975, -0.092 & 52.0 & 52.1 & 0.020 \\
HD\,197818 & 309.416, 0.761 & 43.0 & 35.1 & 0.017 \\
HD\,197212 & 307.980, -1.257 & 47.0 & & 0.120 \\
HD\,198902 & 310.645, -1.486 & 53.0 & 66.8 & 0.076 \\
HD\,196617 & 307.839, 1.809 & 56.0 & 55.7 & 0.035 \\
HD\,123453 & 214.269, 0.015 & 47.0 & 62.3 & 0.018 \\
HD\,123829 & 214.890, -0.206 & 73.0 & &0.022 \\
\hline
\end{tabular}
\end{center}
\end{table*}

For each cloud region we show a plot of the color excesses versus distance
(Figure\ 5, 7, and 9) as well as a plot of the location of the stars 
(Figure\ 6, 8, and 10) in 
relation to the cloud shadows. We use different symbols to discriminate between
reddened stars (filled symbols) and almost unreddened stars (open symbols).
A threshold of E$_{b-y} = 0.02$ has been chosen which is related to an
optical depth exceeding unity in the DS Lex/B filter band. According to the 
relation $N_{\rm H{\sc i}}(atoms/cm^2) = (7.5 \pm 1.0) \times 10^{21} 
{\rm E}_{b-y} - (8.7 \pm 0.55) \times 10^{19}$ (\cite{knude78}), a value of 
E$_{b-y} = 0.02$ is
equivalent to an interstellar H{\sc i} absorption column density of 
$6.3 \times 10^{19} atoms/cm^2$. In principle, the distance of the
EUV-shadowing clouds can be constrained by finding the most distant unreddened
star in front of a cloud and the closest reddened star behind this
cloud. This method can also be applied to determine the distance where optical
depth reaches unity in the regions immediately adjacent to the shadowing cloud
which we need for our new differential cloud technique 
(see Sect.~\ref{diffcloud}).

\begin{figure}
\plotone{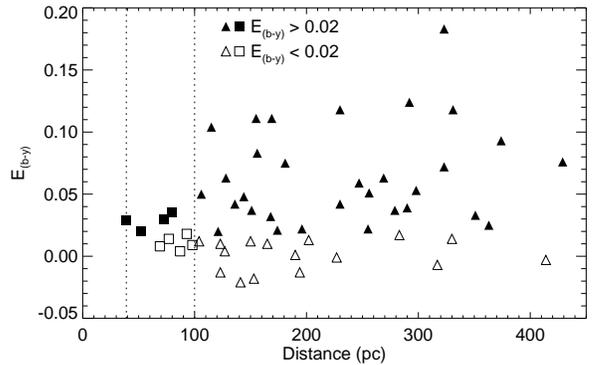}
\caption{\label{165red} Color excesses E$_{b-y}$ versus distances for stars
near lb165-32; triangles and squares belong to distances D $>$ 100\,pc and
D $<$ 100\,pc, open and filled symbols represent excesses 
E$_{b-y} <$ 0.02\,mag or E$_{b-y} >$ 0.02\,mag, respectively. Dotted lines 
indicate the distances of the two clouds (see Sect.~\ref{optobs}).}
\end{figure}

\begin{figure}
\plotone{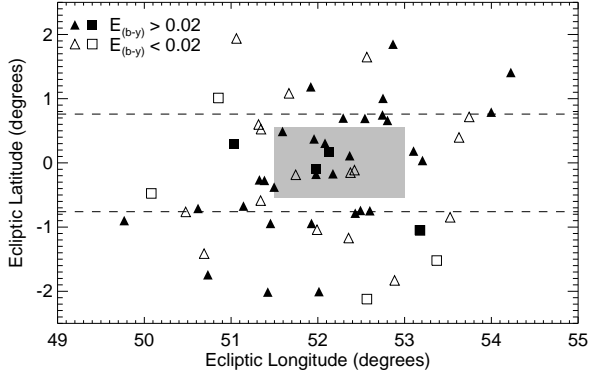}
\caption{\label{165loc} Location of stars shown in Figure~\ref{165red}; same
symbols used as in Figure~\ref{165red}. The shaded area belongs to the FWHM
of the IR-emission / EUV-absorption feature in Figure~\ref{165euv} and the
FWHM of the EUVE scan perpendicular to the Galactic plane, the dashed lines
show the sky area covered by the EUVE observations.}
\end{figure}

In Figure~\ref{165red} we show a plot of the color excesses E$_{b-y}$ versus 
stellar distances for all of our program stars in the region of lb165-32.
Figure~\ref{165loc} shows the positions of our program stars and the location
of the EUV-absorbing / IR-emitting cloud in the scan path of the EUVE 
satellite along the ecliptic plane; two horizontal lines at $\pm 0.76\arcdeg$
indicate the width of the field of view of the DS instrument that has been
used to obtain the EUV background. Of the $\sim$150 stars observed in the 
lb165-32 region, 57 met the criteria required to establish distances using 
the uvby$\beta$ system.
As can be seen from Figure~\ref{165red}, 4 stars at distances less than 
100\,pc (filled squares in Figure~\ref{165red}) show reddening larger than 
E$_{b-y} > 0.02$. Together with the IRAS 100\,$\mu$m sky map, the data in
Figure~\ref{165loc} confirm that the reddening is caused by the cirrus cloud 
in the ISM that is also responsible for the EUV-absorption; the shaded area 
in Figure~\ref{165loc} represents the FWHM area defined by the width
of the IR-emission / EUV-absorption feature (cf. Figure~\ref{165euv}) in 
the direction of the ecliptic plane and the FWHM of the detector response in
ecliptic latitude direction. Since all 
unreddened stars appear outside the cloud boundary and are also located at 
distances D $>$ 40\,pc, the closest reddened star places a limit to the 
distance of the cloud of D $\la$40\,pc. Most stars located beyond 100\,pc 
(triangles in Figure~\ref{165red}), show reddening values well above 
E$_{b-y} > 0.05$. The location of these stars indicate a second much larger 
cloud behind the nearby EUV-absorbing cloud. This larger, more distant cloud 
limits the emitting length for a local continuum in the diffuse EUV background
emission in this direction. The EUV emission must originate in front of 
100\,pc.

Figure~\ref{165loc} shows three stars of low reddening and distances beyond
100\,pc (open triangles) inside the cloud boundary (shaded area). At the 
position of these stars the IRAS 100\,$\mu$m sky map shows a locally lower 
emission level, thus indicating inhomogeneity in the shadowing cloud. 
Compared to the entire shadow areas these smaller ``holes'' cover less than 
$\approx10\%$ and their influence on the EUV background scan is low.

\begin{figure}
\plotone{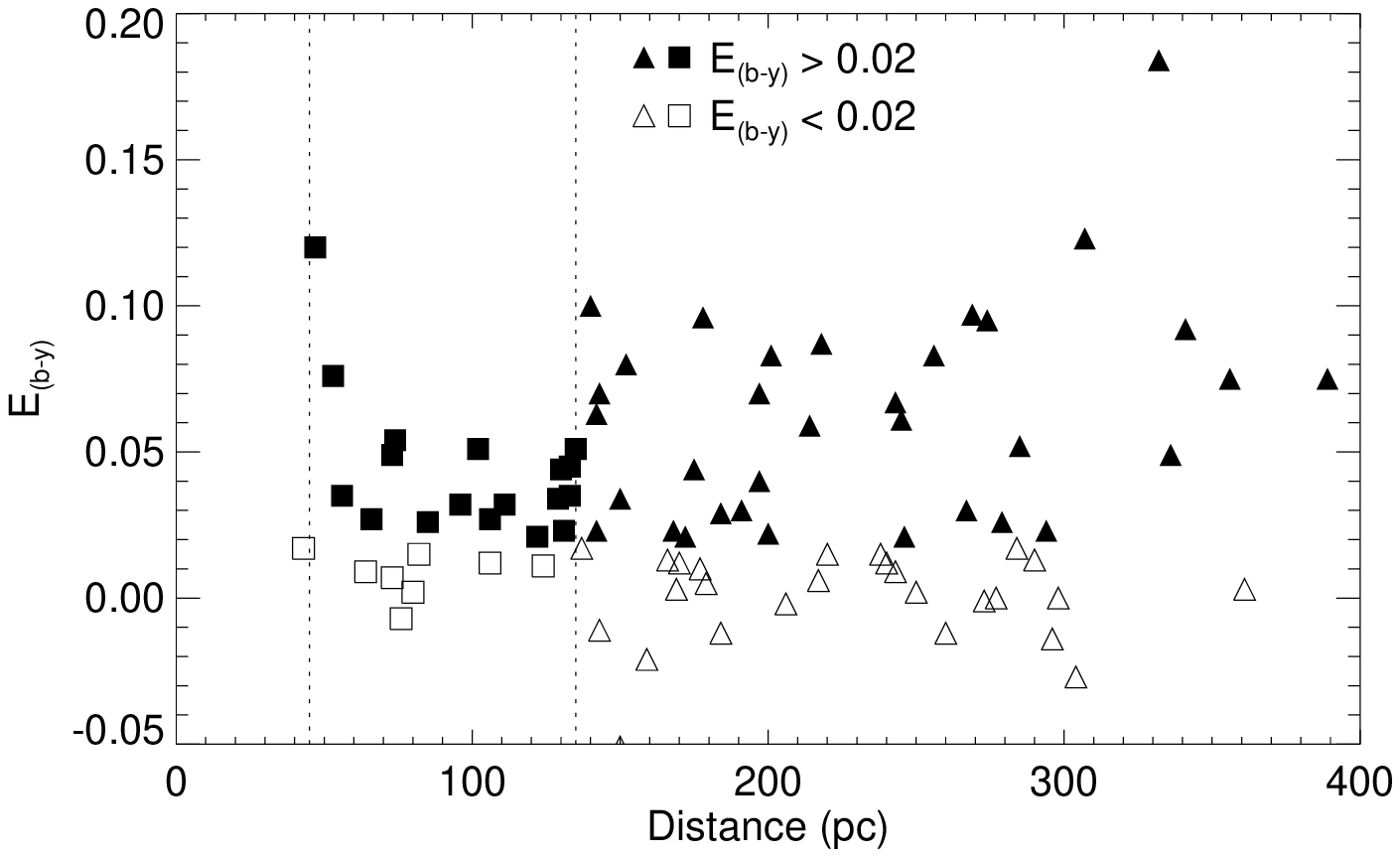}
\caption{\label{27red} Same as Figure~\ref{165red} for the shadow region 
lb27-31; the distance to the background cloud is 125\,pc.}
\end{figure}

\begin{figure}
\plotone{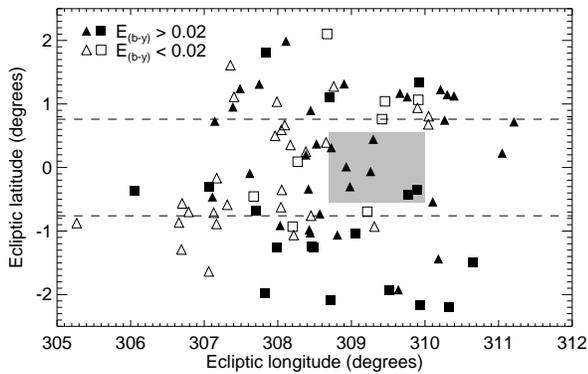}
\caption{\label{27loc} Same as Figure~\ref{165loc} for the shadow region 
lb27-31.}
\end{figure}

In Figure~\ref{27red} we have plotted E$_{b-y}$ against distance for stars in 
the sky region around lb27-31. Figure~\ref{27loc} provides the positions of 
the stars in relation to the cloud shadow; the labeling of the data points is
the same as used in Figure~\ref{165red} and Figure~\ref{165loc}. 
A number of the nearby stars 
(D $<$ 100\,pc) show significant extinction up to E$_{b-y}$ = 0.12\,mag. Two
of these stars are located within the shaded area of Figure~\ref{27loc}
(which marks the most prominent position of the cloud shadow in the EUVE scan).
Together with the closest star (D = 42\,pc) in our sample which shows an 
extinction somewhat below the threshold of E$_{b-y} = 0.02$, the highly 
reddened nearby stars indicate a distance of D $\approx$ 45\,pc for the 
shadowing cloud. We note that the almost unreddened star at D = 42\,pc is 
located at ($l = 309.2\arcdeg, b = -0.7\arcdeg$), right below the 
shaded area in Figure~\ref{27loc}. At D $\approx$ 130 -- 140\,pc there is a 
boundary, beyond this boundary a large number of stars have significant 
extinctions with E$_{b-y} \ge 0.05$mag. This indicates a second cloud 
behind the cloud at D $\approx$ 45\,pc. The distribution of the reddened stars
at larger distances (filled triangles in Figure~\ref{27loc}) confirms the
larger extent of this second cloud. We use a distance of 135\,pc to establish 
the emitting length of the EUV background continuum to this second cloud

\begin{figure}
\plotone{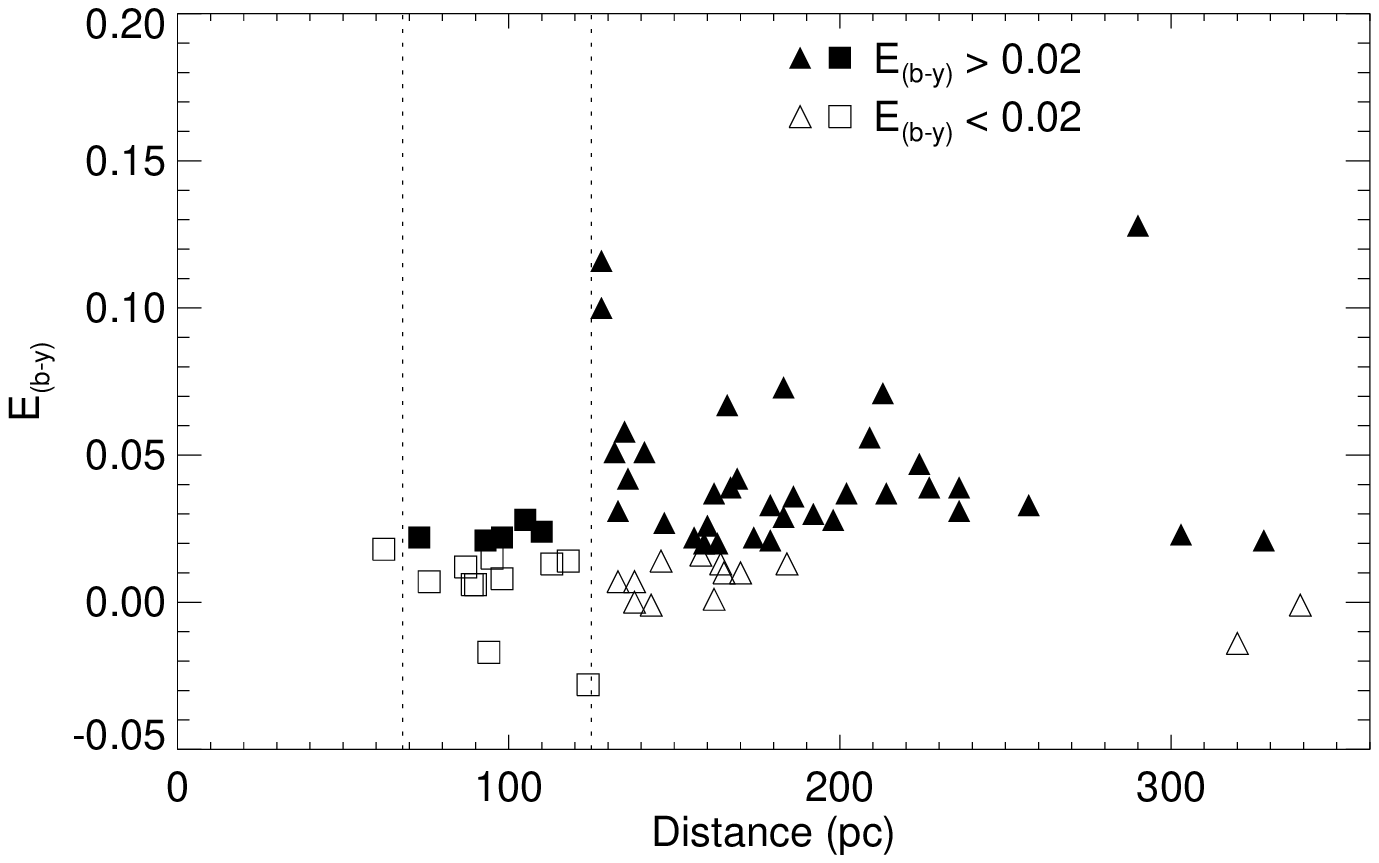}
\caption{\label{329red} Same as Figure~\ref{165red} for the shadow region 
lb329+46; the distance to the background cloud is 135\,pc.}
\end{figure}

\begin{figure}
\plotone{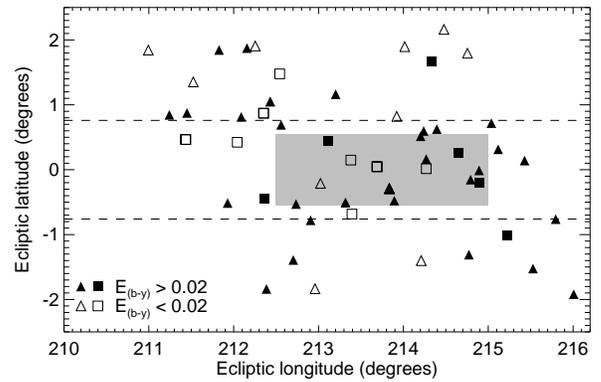}
\caption{\label{329loc} Same as Figure~\ref{165loc} for the shadow region 
lb329+46.}
\end{figure}

The optical data for the third interstellar absorption feature casting a 
shadow in the diffuse EUV background emission (lb329+46) are shown in 
Figures~\ref{329red} and \ref{329loc}. Again, the E$_{b-y}$ versus 
distance plot indicate
the existence of two clouds positioned at distance of D $\approx$ 65\,pc and 
D $\la$ 125\,pc. The positions of the reddened stars at D = 62\,pc 
($l = 214.269\arcdeg, b = 0.015\arcdeg$) and D = 73\,pc ($l = 214.890\arcdeg, 
b = -0.206\arcdeg$) are coincident with the most
prominent absorption feature in the EUV background scan 
(cf. Figure~\ref{329euv}). The somewhat more distant but unreddened stars 
in the EUVE scan region between $l = 213.4 - 214.2\arcdeg$ indicate
less absorbing material in this sky direction which is consistent with
the higher Lex/B count rate in this range. From the
E$_{b-y}$ and distance values of the stars located near ecliptic longitude 
$l = 211\arcdeg$ we determine an upper limit of D $\approx$ 180\,pc
for the distance where optical depths reaches unity (in the EUV).

In all three cloud shadows reported here, our optical data provide 
evidence for a second larger cloud located behind the nearby EUV-absorbing 
cloud. In order to prevent any confusion regarding the different clouds
and the different lines of sight we use the following convention: 
the nearby cloud casting the shadow in the EUV background is designated as the
``shadowing cloud''. The second more distant cloud is denoted as the 
``background cloud''.

\section{Analysis}
\label{result}

\subsection{The differential cloud technique}
\label{diffcloud}

We developed a new method to measure the EUV background flux by means of
EUVE DS observations of nearby clouds in the local ISM, which is independent 
of any zero level calibration of the detector. 
In order to set up this new differential cloud technique the EUVE scan across 
a cloud must cover the shadowing cloud as well as enough regions 
adjacent to the shadowing cloud for a local continuum flux determination in
front of the background cloud. 
We note that this method only works when a smaller cloud is located in front 
of a larger, more distant cloud as in the case presented here for our three 
cloud shadows. Since the EUVE fluxes obtained in the direction of the nearby 
clouds and the more distant background clouds are contaminated by 
the same amount of nonastronomical background, the comparison of the flux at 
the position of the shadowing cloud with the continuum flux originating in 
front of the background cloud provides a differential measurement of the EUV 
background. 

\begin{figure}
\plotone{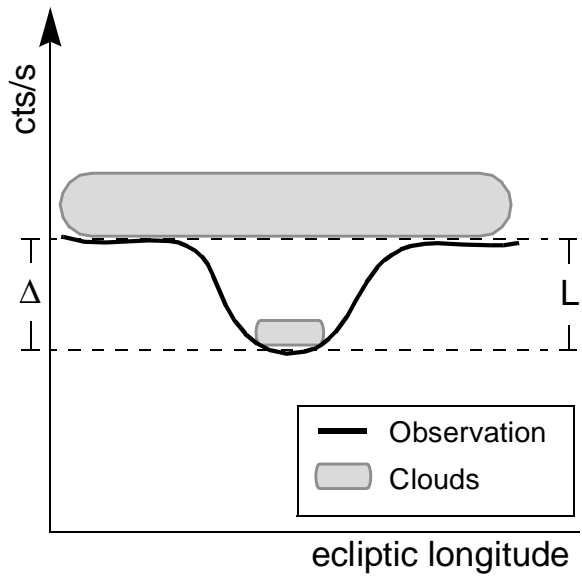}
\caption{\label{dctech} Sketch of our new differential cloud technique. 
A configuration of two clouds is overplotted with the EUV background
scan (solid line) across the smaller nearby shadowing cloud. The local EUV 
background continuum behind the shadowing cloud originates in front of the 
second, larger and more distant background cloud. The observed count rate 
deficit $\Delta$ is associated with the column L between the two clouds in
the ISM along the line of sight.}
\end{figure}

Figure~\ref{dctech} schematically shows our new differential cloud technique.
The observed flux deficit $\Delta$ at the position of the shadowing cloud 
(with respect to the local continuum emission in front of the background cloud)
is associated with a 
specific emitting length L along the line of sight, namely the distance 
between the front of the shadowing cloud and the background cloud.

\subsection{Cloud shadow results}

In Table~\ref{obsval} we provide a summary of the results for our three
cloud shadows. For each cloud, both the shadowing clouds and the background
clouds, we give its distance and the corresponding DS Lex/B filter count rate.
For lb165-32 we use the count rates obtained during the original observation,
so that all the count rates are from the same observation mode
(i.e. the DS). A third measurement given for
lb329+46 is from a location near lb329+46 showing a significantly higher 
EUV flux (cf. Sect.~\ref{euvobs} and Sect.~\ref{optobs}) and a
significantly larger free path.
\begin{table*}
\caption[ ]
{\label{obsval} The observed DS Lex/B count rates and 
the distances for the respective clouds.}
\begin{center}
\begin{tabular}{lcc}
\hline \hline 
cloud & Distance (pc) & Lex/B cts/s\\
\hline
lb165-32 & $\lesssim$40 & $0.35 \pm 0.06$\\
         & $100 \pm 20$&$0.51 \pm 0.06$\\
lb27-31  & $45 \pm 10$&$0.48 \pm 0.03$\\
         & $135 \pm 20$&$0.61 \pm 0.03$\\
lb329+46 & $65 \pm 10$ &$0.36 \pm 0.05$\\
         & $125 \pm 20$&$0.57 \pm 0.05$\\
         & $\lesssim$180 &$0.7 \pm 0.05$\\
\hline
\end{tabular}
\end{center}
\end{table*}

\begin{figure}
\plotone{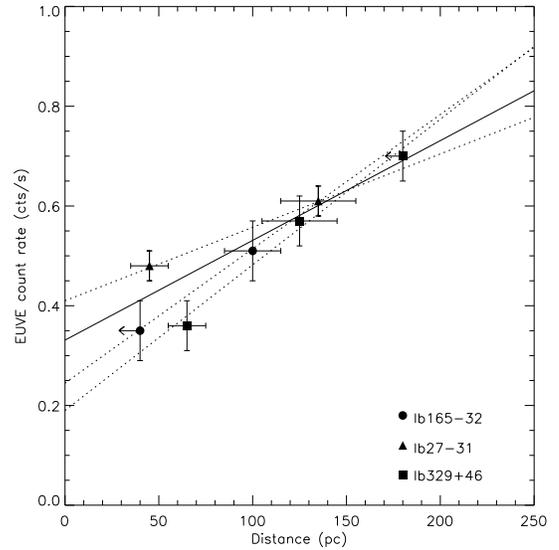}
\caption{\label{ratedist} Plot of the observed count rates versus the 
distance of the respective clouds; distance upper limits are indicated by 
left arrows. The solid straight line provides the results of a parametric
linear regression analysis (slope: 0.0020 $\pm$ 0.0003\,cts/s\,pc$^{-1}$, 
intercept: 0.331 $\pm$ 0.045\,cts/s, see text). The dotted lines show the 
individual results for the different lines of sight (see text).}
\end{figure}

\begin{figure}
\plotone{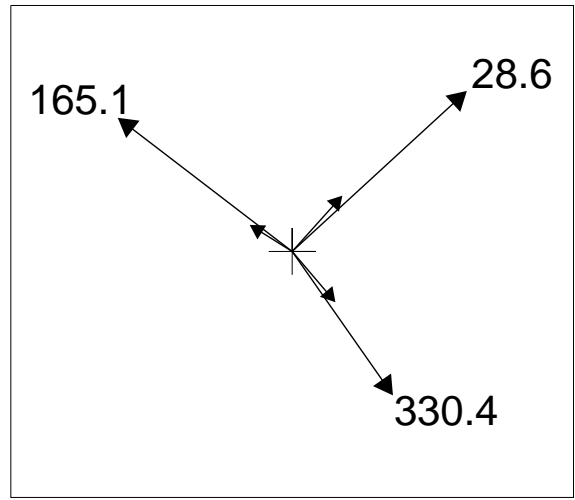}
\caption{\label{clouddir} Projection of the sky directions towards the cloud 
shadows as seen from above the Galactic plane; numbers give the Galactic
longitude, the length of the arrows scales with the projected distance of the 
clouds.}
\end{figure}

In Figure~\ref{ratedist} we plot the DS Lex/B filter count rates versus the 
respective distances for all three cloud shadows. The different cloud shadow
regions are marked by different symbols, the distance upper 
limits are indicated by left arrows. As is evident from Figure~\ref{ratedist} 
almost all the data points are reasonably fit with a straight line. We 
performed 
several correlation tests and confirmed the existence of a tight correlation 
between the DS Lex/B filter count rates and the distances of the clouds.
The formal results of a parametric linear regression analysis of the data 
(including the two distance upper limits) are 
0.002 $\pm$ 0.0003\,cts/s\,pc$^{-1}$ for the slope and 0.331 $\pm$ 0.045\,cts/s
for the intercept of the regression line. This line is plotted in 
Figure~\ref{ratedist}. The slope of the regression line gives the diffuse
EUV flux per emitting length which we use to derive the pressure of the plasma
in the ISM. Figure~\ref{ratedist} includes the regression lines (dotted lines)
for the individual lines of sight. The individual lines of sight show slopes 
between 0.0015\,cts/s\,pc$^{-1}$ (lb27-31) and 0.0027\,cts/s\,pc$^{-1}$ 
(lb165-32) which deviate by -25\% and +35\% from the value obtained by the 
regression analysis including all clouds. We believe these limits reflect 
all uncertainties due to possible variations of the absorption column 
in the foreground of the different clouds, and within these uncertainties
all data of all three lines of sight can be explained by the same slope. 
In the next section, where we compute the pressure of the hot ISM gas, we show 
that these uncertainties can affect the inferred pressure value by only 
10--15\%. The tight correlation between observed diffuse EUV flux and 
emitting length implies essentially the same pressure in all three cloud 
directions. In Figure~\ref{clouddir} we show a plot of the cloud locations
in a projection from above the Galactic plane. As can be seen the detected 
EUV-absorbing clouds are located in three significantly different directions. 
This is important for our conclusions for the hot gas in the ISM and is
further addressed in Sect.~\ref{discus}.

An additional point is that the y-intercept provides a zero level calibration 
for our cloud shadow observations with the EUVE DS instrument. This zero level
can be used to subtract the nonastronomical photons from the EUVE 
observations and allows us to measure absolute fluxes for the diffuse EUV 
background; we note that about 50\% of all detected photons are astronomical
in origin. Consequently, the pressure
value derived from the slope of the regression line not only provides the 
average pressure of the material in the distance range $\approx$ 40 -- 180\,pc
(between the shadowing cloud and the larger background cloud), it also
allows us to determine the thermal pressure of the ISM in the regions in front 
of the three clouds. It is noteworthy that these zero level corrected values 
also give the same pressure for the material in the foreground of the three 
shadowing clouds. We point out that intrinsic absorption along the lines of 
sight can have flattened the apparent correlation between EUV flux and 
distance. Therefore, the given value for the y-intercept represents an upper 
limit.

\subsection{The pressure of the ISM}
\label{pcalc}

We now utilize the results of our cloud shadow observations to 
compute the pressure of the local ISM from the ideal gas law
\begin{equation}
{\rm P}/k = 1.92 \cdot \sqrt{{\rm EM/L}} \cdot {\rm T}
\end{equation}
where P is the pressure, $k$ is the Boltzmann's constant, EM/L is the 
emission measure per emitting length, and T is the temperature; the factor of
1.92 accounts for the number of free particles (electrons + ions) in 
the plasma. The conversion of the observed EUVE DS count rates per emitting 
length into EM/L is model 
dependent. A detailed discussion of the implications of different plasma 
models and plasma emissivity codes on the pressure of the local ISM will be 
given elsewhere.
Here we use the typically assumed model spectrum of a plasma in collisional 
ionization equilibrium at a temperature of $10^6$\,K. To account for 
absorption in the foreground of the clouds we include a foreground 
ISM absorption column of N$_{\rm H} = 5 \times 10^{18}\,{\rm cm}^{-2}$ in our 
model. With the plasma emissivity code provided by 
Landini~\&~Monsignori-Fossi~(\cite{lamofo90}) we derive 
EM/L = 7.3 $\times 10^{-5} {\rm cm}^{-6}$ and a pressure 
P/$k \approx 16500 {\rm cm}^{-3}$\,K; for a Raymond~\&~Smith~(\cite{raym77}) 
model of the same parameters we obtain EM/L = $5.1 \times 10^{-5} 
{\rm cm}^{-6}$ and P/$k \approx 13500 {\rm cm}^{-3}$\,K. 
The column density of the local cloud around our Sun in the three view 
directions varies between 
N$_{\rm H} = 5 \times 10^{17}\,{\rm cm}^{-2}$ (lb27-31) and
N$_{\rm H} = 3 \times 10^{18}\,{\rm cm}^{-2}$ (lb165-32) (P. Jelinsky 1998, 
private communication). This might explain
some of the deviation between the data obtained in the individual view 
directions. However, a 10 times lower absorption column density affects the 
inferred emission measure by 30\% and thus results in an uncertainty of the 
pressure value of only 15\%. 
Any significant absorption by cold gas in front of the clouds exceeding
N$_{\rm H} = 5 \times 10^{18}\,{\rm cm}^{-2}$ would depress the apparent 
diffuse EUV flux with increasing emitting length in the ISM and, thus, 
increase the pressure of the hot gas in the ISM.

We point out that the variation in the
pressure value resulting from the use of different plasma emissivity codes 
(which is a general problem in high energy astrophysics) exceeds the
error in the determination of the EUVE DS count rate per emitting length 
as derived from the slope of the regression line in Figure~\ref{ratedist}. 
The latter results in an error of 10--15\% for the pressure value. Strictly
speaking, our value represents a lower limit for the actual value of the 
pressure. Because we measure the count rate deficit at the
position of a shadowing cloud against the local background, any flux
contribution from behind the shadowing clouds reduces the respective
count rate deficit. Since we cannot definitely exclude such contributions,
the actual value for the slope of the regression line can be higher. 
Additional intrinsic absorption along the lines of sight strengthens this
effect. Therefore, the derived value P/$k = 16500 {\rm cm}^{-3}$\,K provides
a lower limit for the pressure of the local ISM. 

\section{Summary and Conclusions}
\label{discus}

We have presented a re-observation of a cloud shadow in the
diffuse EUV background discovered by Bowyer~et~al.~(\cite{bow95}) and present 
two additional shadows detected in the EUVE DS.

We have developed a new method to derive the diffuse astrophysical EUV
background from the EUVE DS data. This differential cloud technique can be 
applied to any cloud shadow data without a known zero level calibration. 
In our case we can extrapolate the results to the origin and obtain 
a zero level calibration for the DS detector.
This zero level calibration indicates that about 50\% of the detected photons 
are astronomical in origin. More importantly, this zero level calibration 
allows a determination of the diffuse EUV background in the direction of the 
three nearby cloud shadows and hence to derive the pressure in the regions
between us and these clouds. The pressure obtained for these nearby regions is
consistent with the value obtained from the differential measurements for the 
more distant regions along the lines of sight.

The results of our cloud shadow observations with EUVE provide evidence for
a constant pressure in three different directions in the local ISM. 
We derived a pressure of P/$k \approx 16500 {\rm cm}^{-3}$\,K. 
The canonical value obtained from the analysis of solar He~{\sc i} 
584\,\AA~~radiation resonantly scattered by helium in the inflowing cloud is
P/$k = 730 \pm 30 {\rm cm}^{-3}$\,K (e.g., \cite{frisch95}). Based on line 
measurements with EUVE, Vallerga (\cite{vallerga96}) derived a pressure for 
the solar cloud in the range P/$k = 1700 - 2300 {\rm cm}^{-3}$\,K which is in 
good agreement with results obtained by \cite{bertetal85}, P/$k \approx 2600 
{\rm cm}^{-3}$\,K). A comparison of these results shows the pressure of the 
local ISM exceeds the pressure of the cloud surrounding our Sun by a 
factor of $\ge 8$. The original observation (Bowyer~et~al.~\cite{bow95}) of a
pressure imbalance between the solar cloud and the surrounding local ISM
could have been the result of a nearby shock wave which had not yet reached the
Sun. The results reported here are obtained in three different directions.
The isotropy in the results for all these directions is not consistent with 
this hypothesis. 

Flux in the soft X-ray bandpass requires a larger absorption column density 
to reach unit optical depth and it is difficult or impossible to separate the 
locally produced diffuse soft 
X-ray emission from contributions originating at larger distances. Soft X-ray 
observations, therefore, typically provide only an upper limit for the 
pressure of the local ISM. As discussed previously, the pressure obtained from
our cloud shadow observations in the EUV is strictly interpreted as a lower 
limit. Although different plasma codes have been used by different authors, the
pressure derived from our cloud shadow observations with EUVE is about the same
as values derived from soft X-ray observations (e.g., \cite{snowden97}, 
\cite{freyberg97})when we use the same code (P/$k$(EUV) = 13500 cm$^{-3}$K; 
P/$k$(soft X-ray) = 14000 cm$^{-3}$K). The near equality of these
upper and lower limits implies we have measured the true pressure.

Beyond the cloud surrounding our Sun we found no significant intrinsic cold 
gas absorption in the local ISM. This supports the idea of a cavity primarily 
filled with an ionized
plasma. Interstellar absorption lines in the spectra of nearby 
stars have indicated the existence of cloudlets of cold gas (like the local 
cloud around our Sun) in the solar vicinity (Lallement~\cite{lallement96}). 
However, the effect of these cloudlets on the diffuse EUV emission observed in 
the directions of the cloud shadows discussed here is negligible.
We point out that the different line of sights for the six clouds discussed
here show that the local ISM cannot be simply described as a sphere with an 
almost constant radius (\cite{snowcox90}).

Our observations provide evidence for a large pressure imbalance
in the local ISM compared to the local cloud surrounding our Sun. This 
contradicts the basic assumption of almost all available models for the
local ISM of a pressure equilibrium (e.g., \cite{mckost77}, \cite{coxrey87}). 
It is unclear what physical mechanism can maintain such a large 
pressure imbalance. We emphasize that these models as well as the calculations
in the here presented work assume a hot plasma in collisional ionization
equilibrium. A completely different approach is provided by Breitschwerdt \& 
Schmutzler (\cite{breischm94}). Their model calculations are based on an
adiabatically cooling non-equilibrium plasma.
The advantage of this model is that it can at least qualitatively explain many
of the observed features of the local ISM and it does not require pressure 
equilibrium in the local ISM.

\acknowledgments
We thank M. Lampton and J. Vallerga for useful discussions.
T.W.B. acknowledges the support from the Alexander-von-Humboldt-Stiftung (AvH)
by a Feodor-Lynen Fellowship. This work has been supported by NASA contract
NAS 5-30180. R.L. is supported by NASA grant 5-34378 awarded to UAH. J.K. 
thanks the European Southern Observatory for observing time and travel support.

\end{document}